\documentclass[aps,prb,twocolumn,amsmath,amssymb,groupedaddress,floatfix,longbibliography]{revtex4-2}
\setcitestyle{square,numbers}

\usepackage{graphicx}
\usepackage{bm}
\usepackage{color}
\usepackage{epstopdf}

\usepackage[plainpages=false,pdfpagelabels,colorlinks=true,linkcolor=red,urlcolor=blue,citecolor=blue,pdftitle={Title},pdfauthor={},pdfdisplaydoctitle=true,pdfduplex=DuplexFlipLongEdge]{hyperref}

\newcommand{\changed}[1]{#1} 

\begin{document}

\title{Influence of oxygen orbitals and boundary conditions on the pairing behavior in the Emery model for doped ladders}
\author{G\"{o}kmen Polat}
\author{Eric Jeckelmann}
\affiliation{Leibniz Universit\"{a}t Hannover, Institute of Theoretical Physics,  Appelstr.~2, 30167 Hanover, Germany}

\date{\today}

\begin{abstract}
We investigate the Emery model on several ladder-like lattices including two legs of
copper $d$-orbitals and various numbers of oxygen $p$-orbitals.
Pair binding energy, pair spatial structure, density distribution, and pairing
correlation functions are calculated 
using the density matrix renormalization group (DMRG).
We show that a Luther-Emery phase with enhanced pairing correlations can be found for hole doping as well as for electron doping with realistic model parameters.
Ladder properties depend sensitively on model parameters, the oxygen $p$-orbitals taken into account, and boundary conditions.
The pair binding energy  is a more reliable quantity than correlation functions for \changed{ascertaining the occurrence of} pairing in ladders.
Overall, our results for two-leg Emery ladders support the possibility of superconductivity in the hole-doped 2D model. 
The issue is rather to determine which of the various ladder structures and model parameters are appropriate to approximate
the two-dimensional cuprates.

\end{abstract}

\maketitle

\section{Introduction} 

Since the discovery of high-temperature superconductivity in copper oxide compounds, 
correlated electron models on two-dimensional (2D) lattices, such as the Hubbard model,
have been studied to explain a number of properties observed experimentally in these materials~\cite{Dagotto1994,Scalapino2012}. 
These models have also been investigated on one-dimensional (1D) ladder lattices
for almost 30 years with the aim of shedding light on the origin of the unconventional superconductivity~\cite{Dagotto1992,Noack1994}. 
The main motivation is that these ladder models
can be studied with well-established analytical and numerical methods for 1D correlated systems
but may give some insight into the physics of their parent 2D correlated systems~\cite{Dagotto1996}.
Moreover, some cuprate compounds with ladder-like lattice structures
exhibit properties similar to the  layered cuprates.
Undoped materials are found to be charge-transfer insulators~\cite{Azuma1994,Hiroi1995}
while superconductivity has been observed 
upon doping under high pressure~\cite{Uehara1996,Isobe1998,Mayaffre1998}.

The Emery model (also called the three-band Hubbard model)~\cite{Emery1987a,Dagotto1994} was investigated on ladder lattices
two decades ago~\cite{Jeckelmann1998c,Nishimoto2002b,Nishimoto2009} using the density-matrix renormalization
group (DMRG) method~\cite{White1992b,White1993a,Schollwoeck2005,Jeckelmann2008a,Schollwoeck2011}.
These investigations revealed that electron-doped and hole-doped ladders  exhibit different
pairing properties. Nonetheless, bound pairs were found for small doping and a range of physically relevant parameters
in both cases. Moreover, it was observed that the oxygen sites had a strong influence on some ladder properties.
At the time computational resources did not allow for a systematic study of these systems, however.
In particular, the long-range behavior of pairing correlations could not be determined systematically.
In addition, these complex  ladder systems have so far resisted the field-theoretical approaches that have been applied
so successfully  for 1D correlated quantum systems~\cite{Giamarchi2003}.

Recently,  pairing correlations have been investigated for long two-leg Emery ladders
in several works~\cite{Song2021,Song2023,Jiang2023a,Jiang2023b,Yang2024}.
For realistic parameters for superconducting cuprates, 
a Luther-Emery phase~\cite{Luther1974,Giamarchi2003} with dominant pairing correlations is observed
upon electron doping as found in lightly doped Hubbard ladders.
For hole doping, however, these works find no similar phase. Instead
the ladder systems seem to be in phases with dominant density correlations or in an electronic Luttinger liquid phase~\cite{Haldane1981,Giamarchi2003}.

In this article we study these Emery ladders focusing on the pair binding energy,
the role of oxygen orbitals, and the influence of boundary conditions.  We show that 
a Luther-Emery phase can be found for hole doping as well as for electron doping
with realistic model parameters. However,
the ladder properties depend sensitively on these parameters and the boundary
conditions. We find that  the Luther-Emery phase competes with an electronic Luttinger liquid phase
while pairing correlations coexist with charge density correlations, as predicted
by the generic field theory for two-leg electronic ladders.
Therefore, contrary to the recent works~\cite{Song2021,Song2023,Jiang2023a,Jiang2023b,Yang2024}, we conclude that
results for two-leg ladders support the possibility of superconductivity in the hole-doped Emery model on a 2D lattice.
The issue is rather to determine which two-leg ladder structures and model parameters are appropriate
to represent this 2D lattice.

The paper is organized as follows. 
In the second section we introduce the Emery model, the  investigated ladder structures, and the methods used to determine their properties.
We also summarize the predictions of field theory for two-leg electronic ladders.
Results for the Emery model on several ladder structures are discussed in the third section. Finally, we present our conclusion in the fourth section.

\section{Models and methods} 

\subsection{Hamiltonian and ladder structures}

The Emery model~\cite{Emery1987a,Dagotto1994} is a generalization of the Hubbard model
describing holes in the relevant orbitals of a 2D CuO$_2$ lattice.
It includes one $d_{x^2-y^2}$ orbital for each copper atom and one $p$-orbital for each oxygen atom.
The Hamiltonian is 
\begin{eqnarray}\label{eq:hamiltonian}
H=&-& t_{dpx} \sum_{\left\langle ij  \right\rangle, \sigma } \left( p^{\dagger}_{xi\sigma} d^{\phantom{\dagger}}_{j\sigma} + \text{H.c.}\right)   \nonumber \\
&-& t_{dpy} \sum_{\left\langle ij  \right\rangle, \sigma } \left( p^{\dagger}_{yi\sigma} d^{\phantom{\dagger}}_{j\sigma} + \text{H.c.}\right) \nonumber \\
&-& t_{pp} \sum_{\left\langle ij \right\rangle, \sigma} \left( p^{\dagger}_{xi\sigma} p^{\phantom{\dagger}}_{yj\sigma} + \text{H.c.} \right)  \\
&+& \varepsilon_d \sum_{i,\sigma} n^d_{i\sigma} + \varepsilon_p \sum_{\alpha,i,\sigma} n^p_{\alpha i \sigma} \nonumber \\
&+& U_d \sum_i n^d_{i \uparrow} n^d_{i \downarrow} + U_p \sum_{\alpha,i} n^p_{\alpha i \uparrow} n^p_{\alpha i \downarrow} .  \nonumber 
\end{eqnarray}
where $d^{\phantom{\dagger}}_{i\sigma}$ and $d^{\dagger}_{i\sigma}$ are annihilation and creation operators for a hole with spin $\sigma (= \uparrow, \downarrow)$ in the $d$-orbital on 
the copper atom $i$
while $p^{\phantom{\dagger}}_{\alpha i\sigma}$ and $p^{\dagger}_{\alpha i\sigma}$ are the corresponding operators 
for a hole with spin $\sigma$ in the $p_{\alpha}$-orbital ($\alpha=x,y)$ on the oxygen atom $i$.
The hole number operators are denoted $n^d_{i\sigma} = d_{i\sigma}^{\dagger} d_{i\sigma}^{\phantom{\dagger}}$ 
and $n^p_{\alpha i\sigma} = p_{\alpha i\sigma}^{\dagger} p_{\alpha i\sigma}^{\phantom{\dagger}}$.
The first sum in~(\ref{eq:hamiltonian}) runs over all pairs $\langle ij  \rangle$ made of nearest-neighbor $d$-$p_x$ orbitals while 
the second sum is  over all nearest-neighbor $d$-$p_y$ pairs.
The third  sum runs over all nearest-neighbor  $p_x$-$p_y$ orbital pairs.
The four remaining sums run over all $d$ and $p$ orbitals, respectively.

 \begin{figure}
\includegraphics[width=0.45\textwidth]{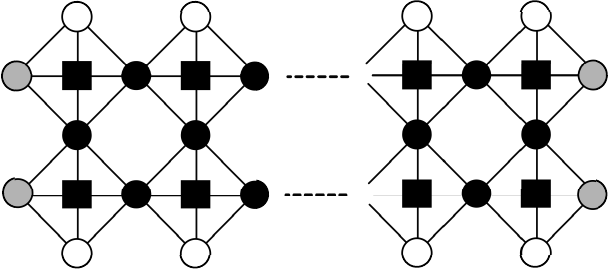}
\caption{\label{fig:structures} Schema of the two-leg ladder lattice structures. Squares and circles represent
copper $d$-orbitals and oxygen $p$-orbitals,  respectively.  Solid black circles show $p$-orbitals included in all lattice structures considered
[$p_x$-orbitals along the (horizontal)  legs  and $p_y$-orbitals on the (vertical) rungs].
Gray circles show the four ladder-end $p_x$-orbitals while open circles show the outer $p_y$-orbitals.
Solid lines represent the  hopping terms between orbitals in the Emery model~(\ref{eq:hamiltonian}).
}
\end{figure}

The Emery model was originally proposed for the isotropic 2D CuO$_2$ lattice.
Here we investigate the Hamiltonian~(\ref{eq:hamiltonian}) on three different ladder-like sublattices.
Our generic ladder structure is sketched in Fig.~\ref{fig:structures}. The chains/legs are in the $x$-direction (horizontal)
while the rungs are in the $y$-direction (vertical).
The first structure includes two legs made of alternating $d$ and $p_x$ orbitals as well as one chain of $p_y$-orbitals making the rungs,
which are represented by solid black squares and circles in the figure. This three-chain ladder Cu$_2$O$_3$ 
was studied in Refs.~\cite{Jeckelmann1998c,Song2021,Song2023}.
The second structure is the five-chain  ladder Cu$_2$O$_5$ including all the orbitals shown
in Fig.~\ref{fig:structures}. It was studied in Refs.~\cite{Nishimoto2002b,Nishimoto2009,Song2021}.
Periodic boundary conditions in the rung direction are used for the third structure
(i.e.,  the $p_y$ orbitals in the top chain are identical to those in the bottom chain in Fig.~\ref{fig:structures}).
The resulting four-chain tube Cu$_2$O$_4$ was studied in Refs.~\cite{Jiang2023a,Jiang2023b,Yang2024}.
Finally, we will also investigate the boundary effects caused by the presence or absence
of the four $p_x$-orbitals at the ladder ends, which are shown in gray in the figure.

All systems studied here include two chains with copper $d$-orbitals. We use the number $L$ of sites with $d$-orbitals in one chain to indicate the system size.
This is also the number of rungs in the ladder in Fig.~\ref{fig:structures}.
Thus each ladder structure contains $2L$ copper sites. The number of sites with oxygen $p$-orbitals is  $3L+b$ for the three-chain ladder Cu$_2$O$_3$,
$5L+b$ for the five-chain ladder Cu$_2$O$_5$, and $4L+b$ for the four-chain tube Cu$_2$O$_4$. The value of $b$ depends 
on the choice for the ladder boundaries: $b=+2$ if the ladder-end $p_x$-orbitals are included, $b=-2$ if they are not.

Undoped cuprate compounds correspond to one hole per copper , i.e. to $2L$ holes. The dopant concentration is
defined relative to the number of $d$-orbitals as $\delta = N/(2L)-1$ where $N$ is the number of holes in the system.
Electron doping corresponds to $N<2L$ ($\delta <0$) while hole doping corresponds to $N>2L$ ($\delta > 0$).

The Hamiltonian~(\ref{eq:hamiltonian}) for a 2D isotropic lattice contains 6 parameters:
 the hopping term $t_{dp} = t_{dpx} = t_{dpy}$ between nearest-neighbor $p$ and $d$ orbitals, the hopping term $t_{pp}$  between nearest-neighbor $p$-orbital pairs,
 the hole site energies $\varepsilon_d$ and $\varepsilon_p$ as well as the Hubbard repulsion $U_d$ and $U_p$ 
 between hole pairs in the same  orbital. 
 We have chosen the phases of the orbitals such that the signs of the hopping terms $t_{dpx}$, $t_{dpx}$, and $t_{pp}$ are constant.
Actually, the model properties depend only on the 
difference in the hole site energies $\epsilon = \epsilon_p - \epsilon_d > 0$ and thus we will only report this value.
 Many different parameter values have been suggested for cuprate compounds~\cite{Emery1987a,Ogata1988,Dagotto1994,Martin1996,Sheshadri2023}
 and used for ladders~\cite{Jeckelmann1998c,Nishimoto2002b,Nishimoto2009,Song2021,Song2023,Jiang2023a,Jiang2023b}.
 In this work we have explored a parameter range around the typical values given in the literature.
 
 As the ladder lattices that we study are anisotropic and inhomogeneous, we also consider 
 hopping terms that can be different in the leg direction ($t_{dpx}$) and the rung direction ($t_{dpy}$).
 Moreover, we will vary the hopping terms between the copper sites and the outer oxygen sites (gray and white circles in Fig.~\ref{fig:structures})
 to study the role of $p$-orbitals and boundary effects.
 
\begin{table}
\centering
\begin{tabular}{||c |l |c |l |l |l |c||} 
 \hline
 $\#$  & $\varepsilon$ & $t_{dpx}$ & $t_{dpy}$ & $t_{pp}$ & $U_d$ & $U_p$\\ [0.5ex] 
 \hline\hline
   1 & 3.0 & 1 & 1.0 & 0.5 & 8 & 3\\
   2 & 2.77 & 1 & 1.08 & 0 & 4.62 & 0\\  
 3 & 2.14 & 1 & 0.64 & 0  & 2.14 & 0 \\  
 \hline
\end{tabular}
\caption{Parameters for the Hamiltonian~(\ref{eq:hamiltonian}) used for most results
presented here. The energy unit is set by $t_{dpx}=1$.}
\label{table}
\end{table}

Most results presented here have been obtained with the sets of parameters listed in Table~\ref{table}.
The set \#1 was used in previous studies of the Emery model on 3-chain and 5-chain ladders~\cite{Nishimoto2002b,Song2021,Song2023}.
We selected the set \#2 to maximize the pair binding energy of the 5-chain ladders Cu$_2$O$_5$
while the set \#3 maximizes the pair binding energy of the 4-chain tube Cu$_2$O$_4$.
\changed{To find optimized parameters for a given ladder, we started from typical values found in the literature~\cite{Dagotto1994}
and using the unit $t_{dpx}=1$ we performed a blind search optimization of the pair binding energy (see below) over the other 5 parameters listed in Table~\ref{table}.}

\subsection{Methods}

We use the DMRG method~\cite{White1992b,White1993a,Schollwoeck2005,Jeckelmann2008a,Schollwoeck2011} to compute the ground-state properties of the Emery model (\ref{eq:hamiltonian}).
DMRG is a well-established numerical method for the study of quasi-one-dimensional correlated systems and has been applied  to Hubbard-type ladder models for 3 decades~\cite{Noack1994}.
We always use open boundary conditions in the leg direction as the DMRG method is more accurate in that case than with periodic boundary conditions. 
The number of rungs $L$ is always even and up to $L=40$. Boundary effects and finite-size effects are significant and will be discussed in sec.~\ref{sec:results}.
The bond dimension is up to $m=9000$  yielding discarded weights of the order of $10^{-8}$ or smaller.
\changed{
We use a noise term to improve the convergence~\cite{White2005}
and we reduce the noise parameter progressively from $10^{-5}$ in the first sweeps with small bond dimensions
to $0$  in the last sweeps with large bond dimensions.}
Most results presented here have been obtained with a DMRG program based on the  C++ version of the ITensor Software Library 
for Tensor Network Calculations~\cite{ITensor}.
We have successfully tested this program with exact diagonalizations~\cite{Weisse2008} of small clusters 
and with another DMRG code that was used in previous works~\cite{Jeckelmann1998c,Nishimoto2002b}.

In this work we focus on four quantities to understand the pairing properties of the Emery model~(\ref{eq:hamiltonian}) on
the ladders shown in Fig.~\ref{fig:structures}: pair binding energy, pair spatial structure, pairing correlations, and density variations.
The most important observable in our analysis is the pair binding energy (PBE).  
Let $E_{0}\left( N_{\uparrow}, N_{\downarrow} \right)$ be the ground-state energy of the ladder system with $N_{\sigma}$ holes of spin $\sigma$.
The PBE is defined as
\begin{eqnarray}\label{eq:Epb}
    E_{pb} &=& 2E_0\left( N_{\uparrow} \pm 1,N_{\downarrow} \right) - E_0\left( N_{\uparrow} \pm 1,N_{\downarrow} \pm 1 \right) \nonumber \\
    &-& E_0 \left( N_{\uparrow},N_{\downarrow} \right) 
\end{eqnarray}
where the plus sign is used for hole doping ($N=N_{\uparrow}+N_{\downarrow} \geq 2L)$ and the minus sign for electron doping
($N\leq 2L)$.
\changed{
The quantity $-E_{pb}$ can be seen as the effective interaction energy between two doped particles (two holes or two electrons, respectively), which generally decreases when the particles become less confined.
Thus we expect the PBE to increase with the ladder length.  This behavior is observed in Hubbard ladders~\cite{Abdelwahab2023}.
In the thermodynamic limit there are 3 possibilities.
First, a positive PBE means that two doped particles  gain energy by building a bound pair.
Second, both particles are unbound and delocalized while the PBE vanishes. Typically, $E_{pb} \leq 0$
in finite ladders in this case.
Third, the PBE can be negative when both particles repulse each other but are trapped by other forces in a finite region, e.g. like two electrons in the helium atom.
Thus  $E_{pb}>0$ is an indication for a 1D precursor of a superconducting phase in higher dimensions
while $E_{pb} \leq 0$ usually signals the absence of pairing.}
In addition, we have calculated spin and charge gaps~\cite{Jeckelmann1998c}, mostly to check for consistency with field-theoretical predictions.

To determine the spatial structure of a pair we calculate
\begin{equation}
S(i,j) = \langle 2 \vert C_{i\uparrow} C_{j\downarrow}   \vert 0 \rangle
\end{equation}
where $\vert 0 \rangle$ is the ground state of an undoped ladder and $\vert 2 \rangle$ denote
the ground state of a ladder with two doped particles (holes or electrons). For hole doping the operator 
$C_{j\sigma}$ creates one hole with spin $\sigma$ in the orbital $j$ while for electron doping
it annihilates the same hole. The indices $i$ and $j$ indicate any of the 
$d$ and $p$ orbitals. 
The normalized distribution
\begin{eqnarray}\label{eq:pair_structure}
\bar{S}_i(j)  = \frac{\vert S(i,j)\vert^2}{\sum_j  \vert S(i,j)\vert^2}.
\end{eqnarray}
 reveals
the spatial distribution of a doped pair. More precisely,  $\bar{S}_i(j)$ can be interpreted as the probability of finding one doped particle
at site $j$ when the other one is at site $i$.

Another evidence for pairing is the behavior of long-range correlation functions.
We have calculated  pairing correlation functions that were already analyzed in previous works~\cite{Nishimoto2002b,Song2021,Song2023,Jiang2023a,Jiang2023b,Yang2024}.
They have the generic form
\begin{equation}\label{eq:correlation}
P(r) = \frac{1}{2} \left(  \left\langle  \Delta_x^{\dagger}  \Delta^{\phantom{\dagger}}_{x^{\prime}} \right\rangle + \left\langle  \Delta^{\phantom{\dagger}}_x \Delta_{x^{\prime}}^{\dagger}  \right\rangle  \right) 
\end{equation}
where the expectation value $\langle \dots \rangle$ is calculated for the $N$-hole ground state of the Hamiltonian~(\ref{eq:hamiltonian}),
$\Delta_x^{\dagger}$  ($\Delta^{\phantom{\dagger}}_x$) creates (annihilates) a singlet hole pair around the position $x$ on the lattice,
and $r$ denotes the distance between the positions $x$ and $x^\prime$. 
A typical example is the rung singlet pair 
\begin{equation}\label{eq:correlation_delta}
 \Delta_x^{\dagger} = \frac{1}{2} \left(  d^{\dagger}_{j\uparrow} d^{\dagger}_{j'\downarrow} - d^{\dagger}_{j\downarrow} d^{\dagger}_{j'\uparrow} \right) 
\end{equation}
on the two $d$-orbitals $j=(x,1)$ and $j^{\prime}=(x,2)$ of the same rung $x\in\{1,\dots,L\}$~\cite{Nishimoto2002b}. Similarly, 
one can define creation and annihilation operators for rung, leg or diagonal pairs of $d$ or $p$-orbitals
and compute correlations between two such operators~\cite{Song2021,Song2023,Jiang2023a,Jiang2023b}.
For instance, rung singlet pairs combining $d$ and $p_y$ orbitals correspond to
\begin{equation}\label{eq:correlation_delta2}
 \Delta_x^{\dagger} = \frac{1}{2} \left(  d^{\dagger}_{j\uparrow} p^{\dagger}_{yj'\downarrow} - d^{\dagger}_{j\downarrow} p^{\dagger}_{yj'\uparrow} \right) 
\end{equation}
with $j$ and $j^{\prime}$ on the same rung $x$.
A 1D system cannot exhibit a true longe-range pairing order and thus we expect any correlation function $P(r)$
to vanish for long distances $r$. Nevertheless, one can search for enhanced pairing correlations 
$P(r) \sim r^{-\alpha}$ with $\alpha < 2$. 
However, one should keep in mind that DMRG yields results for finite distances $r$ only and
that 
it is difficult to extract the correct asymptotic behavior from limited numerical data.

Finally, the local hole density $\rho_i = \langle n^c_i \rangle$ on site $i$ (with $c=d,p$) is defined
as the ground-state expectation value of the hole number operators for a given number of particles $N$ in the system.
It is a useful quantity to understand
differences between the various ladder structures upon doping.
As the density is not constant in undoped ladders, we will discuss the density variation upon doping
\begin{equation}
\Delta \rho_i(\delta) = \rho_i(\delta) - \rho_i(\delta=0)
\end{equation}
where  $\rho_i(\delta)$ is the ground-state density for a given doping $\delta$.
For better comparison we will show normalized distributions
\begin{eqnarray}\label{eq:local_density}
\bar{\rho}_i(\delta) = \frac{\Delta \rho_i(\delta)}{\sum_i \vert  \Delta \rho_i(\delta)\vert}.
\end{eqnarray}
When the density variation has the same sign for all orbitals $i$, $\bar{\rho}_i(\delta)$ is just the density variation $\Delta \rho_i(\delta)$ divided by the
number of doped particles $\vert N-2L \vert = \vert\delta\vert 2L$.

\subsection{Field theory \label{sec:ft}}

While most correlated ladder systems are not exactly solvable, 
their generic properties have been largely determined using field-theoretical methods~\cite{Giamarchi2003}.
In this section we summarize some results that will help us understand doped Emery ladders.
Although the field-theoretical approach is based on the weak-coupling limit of
1D lattice models, exact solutions and numerical results have confirmed that
its predictions usually remain qualitatively correct even in the strong-coupling regime.

First of all, it is important to realized that the electronic structure of the ladder model~(\ref{eq:hamiltonian}) without interactions ($U_d=U_p=0$) 
includes one or two 1D bands crossing the Fermi energy in the regime corresponding to 
lightly doped cuprates. This observation applies to 
the three types of ladder structures considered here (three-chain and five-chain ladders as well as  four-chain tube).
The band structures are presented in the appendix.
Thus the lightly doped Emery ladder belongs to the same class of 1D systems as the two-leg Hubbard ladder.
(When we mention Hubbard ladders we always mean the single-orbital Hubbard model on a ladder lattice.)

We focus on the properties of these ladder systems at incommensurate band fillings,
which is relevant for the doped Emery model ($\delta \neq 0$) as well as
for the two-leg Hubbard ladder doped away from half filling.
We have to distinguish two main cases. First, if
only one band crosses the Fermi energy in the non-interacting limit,
the (weakly) interacting system is an electronic Luttinger liquid with one gapless
spin excitation mode and one gapless charge excitation mode.
This is a C1S1 phase in the notation of Ref.~\cite{Balents1996}.
Second, if two  bands cross the Fermi energy in the non-interacting limit,
the (weakly) interacting system is a Luther-Emery liquid with one gapless
charge excitation mode. Spin excitations and the other charge mode are gapped.
This is a C1S0 phase in the notation of Ref.~\cite{Balents1996}.

For instance, these two phases are seen in the two-leg Hubbard ladder. 
In the non-interacting limit its spectrum is made of two bands separated by a gap if  
the rung hopping $t_{\perp} > 0$ is larger than twice the leg hopping $t_{\parallel} > 0$
and two overlapping bands otherwise.
Thus close to half filling the system is in a C1S0 phase for $t_{\perp} \alt 2t_{\parallel}$
but in a C1S1 phase for $t_{\perp} \agt 2t_{\parallel}$~\cite{Balents1996}.

For Emery ladders~(\ref{eq:hamiltonian}) we show in the appendix that
the lowest two bands overlap for $t_{dpy} < \sqrt{2} t_{dpx}$ in the three-chain and five-chain ladders
and  for $t_{dpy} < t_{dpx}$ in the four-chain tube.
Thus a Luther-Emery phase is possible in lightly doped ladders
when the rung hopping is not too large, i.e. for 
$t_{dpy} \alt \sqrt{2} t_{dpx}$ or $t_{dpy} \alt t_{dpx}$, respectively.
For larger $t_{dpy}/t_{dpx}$, the system should be in a Luttinger liquid phase.

A quasi-long-range order indicates an ordered state that a 1D system would like to have
but cannot be stable in one dimension. The long-wave-length  behavior of correlation functions
can be determined using field-theoretical methods. Thus we know the possible quasi-long-range 
orders in a given class of 1D systems. 
In an electronic Luttinger liquid the dominant correlations are spin and charge fluctuations with a wave vector  $2k_F$.
In an electronic Luther-Emery liquid 5 types of quasi-long-range 
order are possible depending on the Luttinger parameter $K$ for the symmetric charge mode
and the back-scattering  strength~\cite{Giamarchi2003} but only two types are relevant for the present study.
Pairing correlations of the $d$-wave type decay as 
\begin{equation}
\label{eq:dwave}
P(r) \sim \frac{1}{r^{1/(2K)}}
\end{equation}
while
charge-density-wave (CDW) correlations with a wave vector $4k_F$ decay as 
\begin{equation}
\label{eq:cdw}
C(r) \sim \frac{1}{r^{2K}}
\end{equation}
for $K < 1$.
Thus pairing correlations are enhanced for $K > 1/4$ and are dominant
for $1/2 < K < 1$ while CDW correlations are enhanced for any $K<1$
and are dominant for $K < 1/2$.
Note that we use the definitions of Ref.~\cite{Giamarchi2003}, which differs
from Refs.~\cite{Dolfi2015,Song2023} by a factor 2.
Moreover, it is helpful to realize that the wave vector $2k_F$ of charge fluctuations
in the Luttinger liquid (i.e., in a single, almost filled or empty band)
is the same as the wave vector $4k_F$ of the CDW correlations in the
Luther-Emery phase (i.e., in two partially filled bands, close to half filling together).
Thus the Luttinger liquid phase and the Luther-Emery phase with dominant CDW fluctuations
can only be distinguished using other properties, e.g. the PBE and the spin gap.

In repulsive lightly-doped two-leg Hubbard ladders, $d$-wave pairing correlations always seem
to dominate, i.e. $1/2 < K < 1$~\cite{Noack1994,Dolfi2015,Shen2023}. Nonetheless, $4k_F$-CDW fluctuations are clearly visible 
in the Friedel oscillations~\cite{Dolfi2015}. 
Moreover, both types of quasi-long-range orders
are found in the C1S0 phase of two-leg $t-J$ ladders~\cite{White2002}.
The  other three types of quasi-long-range order that can occur in an electronic Luther-Emery phase 
do not play a role in the Emery model with realistic parameters.
For instance, orbital current correlations can be dominant in doped ladders
but not for model parameters leading to an insulating ground state in the undoped ladder~\cite{Nishimoto2009}.

\section{Results \label{sec:results}}

\begin{figure}
\includegraphics[width=0.48\textwidth]{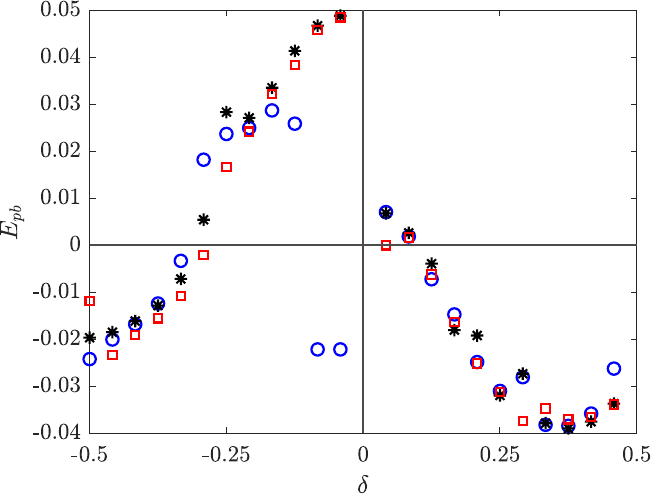}
\caption{\label{fig:pbe31} Pair binding energy  $E_{pb}$ of three-chain ladders with length $L=24$ as a function of the doping $\delta$
for the first parameter set in Table~\ref{table}.  Results are shown for three different boundary configurations:
without ladder-end $p_x$-orbitals (blue circle), with ladder-end $p_x$-orbitals (star), and for strong hybridization with ladder-end $p_x$-orbitals (red square).
}
\end{figure}

First, we summarize some basic results found in previous studies of the Emery ladders~\cite{Jeckelmann1998c,Nishimoto2002b}  and confirmed
in the more recent investigations, including the present one.
For physically relevant parameters the Emery model is a charge-transfer insulator
with holes mostly distributed on the copper sites  at a filling corresponding to undoped cuprate compounds.
Thus this model allows us to study the effects of hole and electron doping on a 
charge-transfer insulator.
In a range of physically relevant parameters and dopings pairing correlations decay as a power-law and have a $d$-wave-like structure in the sense that 
the rung-leg correlations have an opposite sign to the rung-rung or leg-leg correlations.
There are some significant differences between ladders doped with holes and with electrons.
First, doped holes ($\delta > 0$) go primarily onto the oxygen sites while electron doping ($\delta < 0$)
primarily removes holes from the copper sites.
Second, the pairing correlations are stronger for the electron-doped case than for the hole-doped case.
Third, the internal structure of a hole pair differs from that of an electron pair. 

\subsection{Three-chain ladders Cu$_2$O$_3$ \label{sec:3chain}}

We first discuss the properties of three-chain ladders.  Figure~\ref{fig:pbe31} shows the PBE
as a function of the doping $\delta$ for the first parameter set in Table~\ref{table}.
Without ladder-end oxygen orbitals (the gray circles in Fig.~\ref{fig:structures}) a positive PBE
is found for a small range of electron doping. The PBE is negligible or negative for hole doping
as well as for low and high electron doping.  
Pairing correlations are enhanced only in the doping range
where $E_{pb}$ is positive. The strongest correlations are found between $d$-orbital rung pairs  (\ref{eq:correlation_delta}).
These results agree with previous findings, in particular Ref.~\cite{Song2023}.

\begin{figure}[t]
\includegraphics[width=0.48\textwidth]{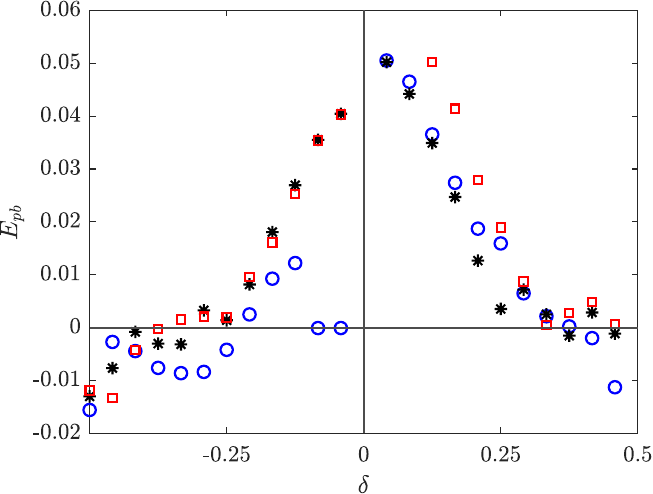}
\caption{\label{fig:pbe32} Pair binding energy  $E_{pb}$ of three-chain ladders with length $L=24$ as a function of the doping $\delta$
for the second parameter set in Table~\ref{table}.  Results are shown for three different boundary configurations:
without ladder-end $p_x$-orbitals (blue circle), with ladder-end $p_x$-orbitals (star), and for strong hybridization with ladder-end $p_x$-orbitals (red square).
The red squares for $0 < \delta < 0.1$ are no visible because they lie too low ($E_{pb} \approx -0.23t_{dpx}$).
}
\end{figure}

Our numerical results for light hole doping are compatible with 
a Luther-Emery phase with a very small \changed{PBE}  and no enhancement of pairing correlations,
which agrees with the field-theoretical predictions in sec.~\ref{sec:ft}, as two bands cross
the Fermi energy in the noninteracting three-chain ladders with these parameters, see the Appendix.
However, we cannot exclude an electronic Luttinger liquid phase, especially where the PBE is negative.
In both cases we observe a static CDW due to a pinning by the ladder boundaries.

\begin{figure}
\includegraphics[width=0.48\textwidth]{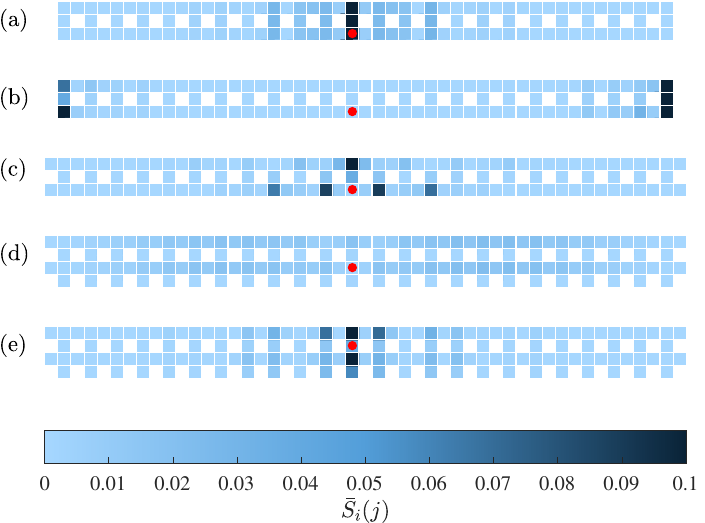}  
\caption{\label{fig:pair_structure} Pair spatial structure $\bar{S}_i(j)$ defined in ~(\ref{eq:pair_structure}). The fixed site $i$ is marked by a red circle.
(a) Hole-doped three-chain ladder.
(b) Electron doped three-chain ladder without oxygen orbitals at its ends.
(c) Electron-doped three-chain ladder with ladder-end oxygen orbitals.
(d) and (e) Hole-doped four-chain tube.
In the first four cases the second parameter set in Table~\ref{table} was used.
In the last case, the third parameter set was used.
The ladder length is $L=24$ in all cases.
}
\end{figure}

However,  this is not a generic outcome  for hole-doped Emery ladders.
Figure~\ref{fig:pbe32} shows the PBE for the second parameter set in Table~\ref{table}.
Without ladder-end oxygen orbitals  we find a clearly positive PBE for a wide range 
of hole doping. The existence of pairs is confirmed by the analysis of the pair spatial structure~(\ref{eq:pair_structure}).
An example is shown in Fig.~\ref{fig:pair_structure}(a). Clearly, the second doped hole is localized close to the first one.
This analysis also reveals the significant weight of central $p_y$-orbitals for the hole pair.

\begin{figure}
\includegraphics[width=0.48\textwidth]{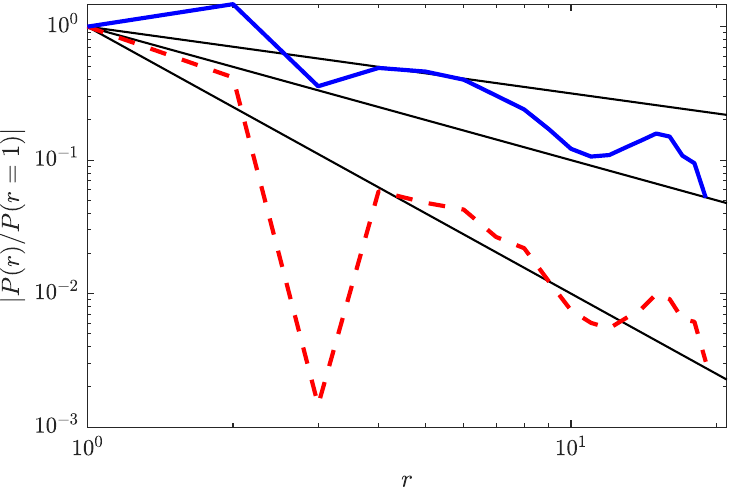}
\caption{\label{fig:correl32h} Correlation functions for $d$--$p_y$ rung pairs (\ref{eq:correlation_delta2}) in three-chain ladders with ladder-end oxygen orbitals,
length $L=40$, and doping $\delta = 0.1$
for the first (red dashed) and second (blue) parameter sets in Table~\ref{table}. 
The straight lines correspond to $r^{-\alpha}$ with $\alpha$ = 1/2,1, and 2.
}
\end{figure}

Pairing correlations are strongly enhanced for hole-doped ladders with the second parameter set.
Contrary to the electron-doped ladder, the strongest correlations are for $d$--$p_y$ rung pairs (\ref{eq:correlation_delta2}).
In fig.~\ref{fig:correl32h} we compare this correlation function for the two parameter sets.
Clearly, pairing correlations decrease as slowly as $1/r$ with the distance $r$ when the PBE is positive (second parameter set) while they decay as fast as in the 
noninteracting limit ($\sim 1/r^2$) when $E_{pb}=0$ (first parameter set).

\begin{figure}[b]
\includegraphics[width=0.48\textwidth]{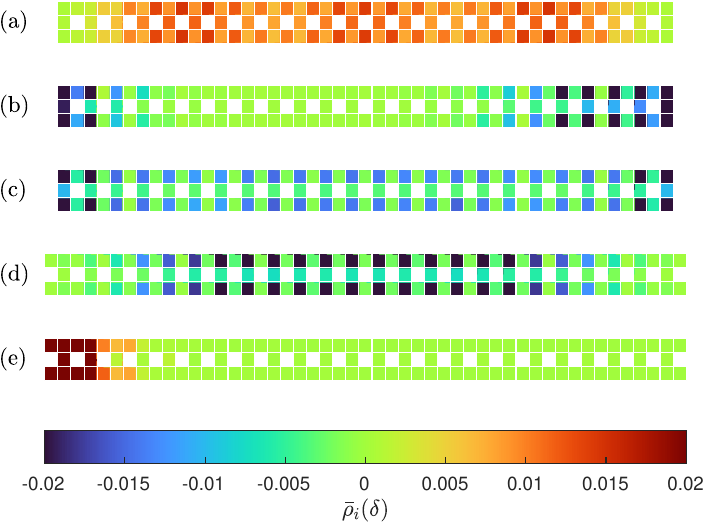}
\caption{\label{fig:density32} Hole density variation $\bar{\rho}_i(\delta)$  of three-chain ladders with length $L=24$ 
and the second parameter set in Table~\ref{table} for various dopings $\delta$.  
The first three plots show ladders without oxygen sites at their ends
while the lower two plots show ladders including these sites.
From top to bottom:
(a) $\delta=+1/8$, (b) two doped electrons $(\delta=-1/24)$, (c) $\delta=-1/8$, (d) two doped electrons $(\delta=-1/24)$,
and (e) two doped holes $(\delta=+1/24)$ with $t^\prime_{dpx}=2t_{dpx}$.
}
\end{figure}

Moreover, we observe a CDW in the hole density variation $\bar{\rho}_i(\delta)$, which is shown in Fig.~\ref{fig:density32}(a).
It can be interpreted as the signature of $4k_F$-CDW fluctuations that are made visible as a static CDW due to a pinning by the ladder boundaries.
Thus our numerical results for hole doping and the second parameter set are compatible with a Luther-Emery phase with 
a Luttinger parameter close to the boundary ($K=1/2$) between dominant $4k_F$-CDW correlations and dominant $d$-wave pairing
(see sec.~\ref{sec:ft}).
Note that we have not attempted to determine the Luttinger parameter $K$ from the decay of correlation functions
or the density oscillations, unlike in recent studies~\cite{Song2021,Song2023,Jiang2023a,Jiang2023b,Yang2024}, 
because we can simulate only relatively short ladders

\begin{figure}
\includegraphics[width=0.48\textwidth]{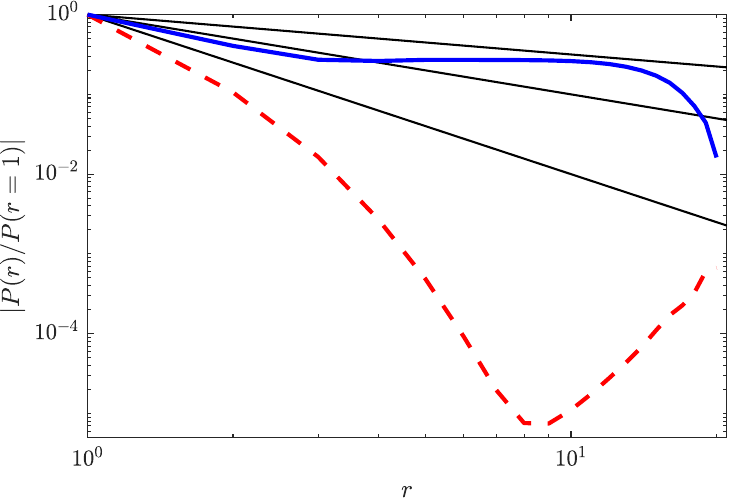}
\caption{\label{fig:correl32e} Correlation functions for $d$-orbital rung pairs (\ref{eq:correlation_delta}) in three-chain ladders with length $L=40$ and $\delta = -0.05$
for the first parameter set in Table~\ref{table}.  The red dashed  curve shows results without ladder-end $p_x$-orbitals
while the solid blue curve shows results with these orbitals.
The straight lines correspond to $r^{-\alpha}$ with $\alpha$ = 1/2,1, and 2. }
\end{figure}

A similitude between both parameter sets is the nonpositive PBE at low electron doping
for ladders without oxygen sites at their ends, as seen in Figs.~\ref{fig:pbe31} and~\ref{fig:pbe32}. 
Pairing correlations also decay
very rapidly in that case, see Fig.~\ref{fig:correl32e}. 
This behavior was also reported in previous works on three-chain ladders~\cite{Jeckelmann1998c,Song2023} but was not found in five-chain ladders~\cite{Nishimoto2002b}.
We have found that this absence of pairing  is just an artefact of the choice of the boundary configuration at the ladder ends. 
Indeed the first couple of doped electrons are trapped at the ladder ends as revealed by their density distribution
in Fig.~\ref{fig:density32}(b) and thus pairing can only occur at higher doping. 
Note that the attraction of doped electrons by the ladder ends remains significant at finite doping, as shown
in Fig.~\ref{fig:density32}(c). This boundary effect also raises questions about the accuracy of the Luttinger parameters $K$
that were calculated assuming the simple Friedel oscillations predicted by field theory in recent 
works~\cite{Song2021,Song2023,Jiang2023a,Jiang2023b,Yang2024}.
Naturally, the localization of doped electrons is a finite-size effect  and we expect
pairing to occur as soon as $\delta < 0$ in the thermodynamic limit. 
Nevertheless, analyzing the role of the boundary sites yields useful information
for reducing finite-size effects in DMRG simulations.

For three-chain ladders including ladder-end oxygen sites (the gray circles in Fig.~\ref{fig:structures}) 
Fig.~\ref{fig:density32}(d) shows that two doped electrons are not trapped at the ladder ends but delocalized
on the lattice.
Correspondingly,  the PBE is positive for all electron doping $-0.25 \alt \delta < 0$, as indicated by stars
in Figs.~\ref{fig:pbe31} and~\ref{fig:pbe32}.
The difference between the two type of boundaries is also illustrated by the pair spatial structure~(\ref{eq:pair_structure})
displayed in Fig.~\ref{fig:pair_structure}. Clearly, the second doped electron is localized at the ladder ends when
they are terminated by $d$ and $p_y$ orbitals in Fig.~\ref{fig:pair_structure}(b) but it is close to the first doped electron 
when the ladder-end sites are $p_x$-orbitals in  Fig.~\ref{fig:pair_structure}(c).
In contrast to the hole pair [Fig.~\ref{fig:pair_structure}(a)],  the electron pair is mostly distributed over $d$-orbitals.
This difference in the internal structure of hole and electron pairs was observed previously 
using complex correlation functions~\cite{Nishimoto2002b}.

Moreover, pairing correlations are now strongly enhanced. For instance, Fig.~\ref{fig:correl32e}
compares the correlation functions for $d$-orbital rung pairs (\ref{eq:correlation_delta}) with and without
the ladder-end oxygen sites.
Thus our numerical results for electron doping $-1/4 \alt \delta < 0$ and both parameter sets fully agree with a Luther-Emery phase
with dominant $d$-wave pairing correlations and weaker CDW fluctuations when the ladder-end oxygen sites are taken into account.
We note that the hole-doped ladders are barely affected by the presence of the additional oxygen sites.

Conversely, one can apply a strong hybridization $t^\prime_{dpx}$ between the ladder-end $p_x$-orbitals and their
nearest-neighbor $d$-orbitals. We show  the PBE obtained with an hybridization twice as large as 
the bulk hybridization ($t^\prime_{dpx}=2t_{dpx}$) in Figs.~\ref{fig:pbe31} and~\ref{fig:pbe32}. Clearly, the results for electron doping
are barely changed but now the PBE becomes negative at low hole doping (up to four doped holes) for the second parameter set
in Fig.~\ref{fig:pbe32}.  The hole density variation $\bar{\rho}_i(\delta)$ reveals that the doped holes are trapped at the ladder ends in that case.
For instance, Fig.~\ref{fig:density32}(e) reveals that the first two doped holes are trapped at one ladder edge.
In  Fig.~\ref{fig:pbe32} we also see that the PBE at intermediate hole doping is larger for $t^\prime_{dpx}=2t_{dpx}$  than
for $t^\prime_{dpx}=t_{dpx}$. This can also be explained as a consequence of the boundary effect. As the first four doped holes are bound at the ladder edges,
the effective doping of the rest of the ladder is $\delta^\prime= (N-4)/(2L)-1$ instead of $\delta =N/(2L)-1$ leading to a shift $2/L \approx 0.08$ of the curve $E_{pb}$ as a function of $\delta$.
In summary, doped electrons are attracted to the ladder ends in the absence of ladder-end oxygen sites or for 
a weak hybridization $t^\prime_{dpx}$ while doped holes are attracted there
for a strong hybridization of these sites with the ladder.

\begin{figure}
\includegraphics[width=0.48\textwidth]{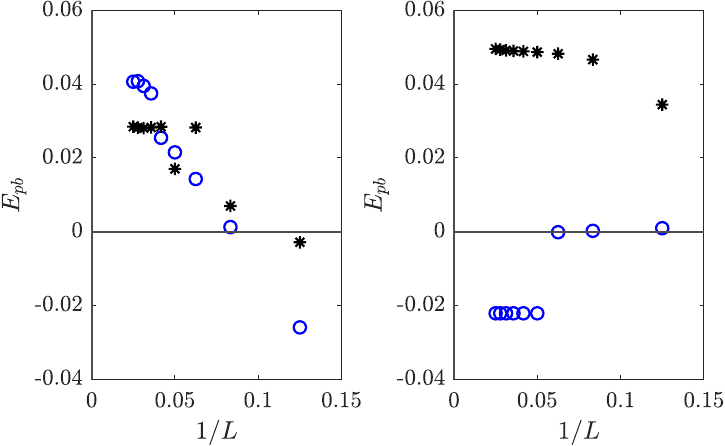}
\caption{\label{fig:scaling}  PBE in three-chain ladders as a function of the inverse ladder length $1/L$  for the first parameter set in Table~\ref{table}.
Blue circles shows the PBE for  ladders without ladder-end oxygen sites while 
black \changed{stars} are results for ladders including these sites.
The left panel shows results for a fixed doping $\delta=-1/4$  and the right panel 
shows results for two doped electrons ($\delta=-1/L$).
}
\end{figure}

As pointed out in~\cite{Song2021,Song2023} there is a strong asymmetry between hole and electron doping 
in three-chain ladders  without ladder-end $p_x$-orbitals for the first parameter set in Table~\ref{table}.
This is illustrated by the PBE as a function of $\delta$ in Fig.~\ref{fig:pbe31}.
With ladder-end $p_x$-orbitals and the second parameter set, however, the behavior of the PBE
is similar for hole and electron doping (see the stars in Fig.~\ref{fig:pbe32}).  

Finite-size effects on the PBE were found to be significant in a previous study~\cite{Jeckelmann1998c}.
The present analysis shows that it was essentially due to this boundary effect.
The influence of the boundary configurations on the finite-size scaling is illustrated in Fig.~\ref{fig:scaling}.
We see that the convergence of the PBE is fast and regular for ladders including
the ladder-end oxygen sites. Our results for $L \geq 24$ already agree with
the PBE calculated using larger ladders in Ref.~\cite{Song2023}.
 Note that a positive PBE usually increases with the ladder length $L$, \changed{as found previously in Hubbard ladders~\cite{Abdelwahab2023}},
 while 
other gaps usually decrease with $L$ in the absence of boundary localization.  Thus a positive PBE in a finite ladder generally indicates 
a positive PBE in the thermodynamic limit.
\changed{
The right panel of Fig.~\ref{fig:scaling} illustrates the effects of the ladder-end O sites at low doping.
In ladders with these sites two doped electrons form a bound pair with a positive PBE. Without these sites both electrons are strongly localized at the ladder ends 
but their mutual repulsion leads to a negative PBE.}

Therefore, the ladder-end configuration affects the properties of doped three-chain Emery ladders.
It is necessary to take into account this issue when discussing the pairing properties of these ladders.
The best choice  for three-chain ladders seems to be the inclusion of ladder-end $p_x$-orbitals 
with the same hybridization $t_{dpx}$ as in the ladder bulk,
i.e. the lattice made of the black and gray symbols in Fig.~\ref{fig:structures}.
With this configuration, we find that all our results are fully compatible with the field-theoretical predictions in Sec.~\ref{sec:ft}.

\subsection{Five-chain ladders Cu$_2$O$_5$ \label{sec:5chain}}

\begin{figure}
\includegraphics[width=0.48\textwidth]{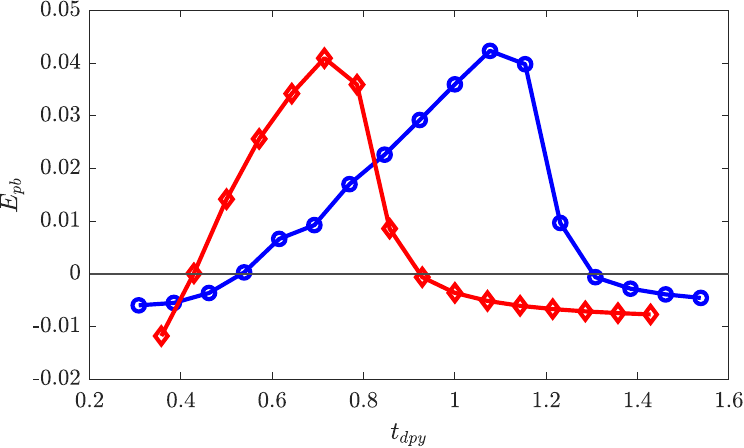}
\caption{\label{fig:pbe52} PBE of Emery ladders with length $L=24$ and two doped holes as a function of the rung hopping $t_{dpy}$.
The blue circles indicate results for a five-chain ladder and the other parameters of the second set in Table~\ref{table}.
The red diamonds show results a four-chain tube and the other parameters of the third set in Table~\ref{table}.
}
\end{figure}

In a previous work it was found that including outer oxygen sites (white circles in Fig.~\ref{fig:structures}) was necessary to obtain 
a qualitatively correct description of the magnetic properties~\cite{Nishimoto2002b}.
The discussion of the three-chain Emery ladder in the previous section has also revealed the importance of the
ladder-end oxygen sites (gray circles in Fig.~\ref{fig:structures}) . Thus we now discuss the properties of five-chain ladders including 
all sites shown in Fig.~\ref{fig:structures}.

Simulating five-chain ladders with DMRG is significantly more difficult than for three-chain ladders.
Consequently,  five-chain Emery ladders have been studied relatively rarely.
Thus we have first investigated the PBE as a function of the Hamiltonian parameters
$U_d > 0$, $U_p \geq 0$, $t_{dpy}$, $t_{pp}$, and $\epsilon > 0$.

In fig.~\ref{fig:pbe52} we show the PBE of a  ladder with two doped holes as a function of the rung hopping $t_{dpy}$
while other parameters are kept fixed. 
$E_{pb}$ is positive in a finite range with a maximum for $t_{dpy} \approx 1.1 t_{dpx}$ but clearly negative for hybridization higher than $t_{dpy} \agt1.4 t_{dpx}$. 
This behavior can be understood from the band structure
of the noninteracting five-chain ladder, which is presented in the appendix.
As the number of 1D bands at the Fermi energy drops from 2 for $t_{dpy} < \sqrt{2} t_{dpx}$
to 1 above this limit, one expects a transition from the Luther-Emery (C1S0) phase to the Luttinger phase (C1S1)
around this critical value (see the discussion of field-theoretical results in sec.~\ref{sec:ft}).
One could object that the 2D Emery model for superconducting cuprate compounds is isotropic ($t_{dpy} = t_{dpx}$).
However, as a ladder lattice obviously breaks the symmetry between the $x$ and $y$ direction, it is possible that
the 2D lattice is better approximated using different hoppings in the leg and rung directions.
\changed{Note that the non-monotonic behavior of the PBE in fig.~\ref{fig:pbe52}  is also observed
in Hubbard ladders~\cite{Abdelwahab2015,Abdelwahab2023}.}

We have found that the local repulsion $U_p > 0$ on the oxygen sites and the hopping between $p$-orbitals $t_{pp}$
are unfavorable for pairing in hole doped ladders. When the PBE is positive, it is the largest at $U_p=0$ and $t_{pp}=0$,
but there is a small range of parameters $U_p>0$ and $t_{pp}\neq0$ around the maximum where $E_{pb}$ remains positive.
Outside this range these two interactions push 
the hole-doped ladder toward the Luttinger liquid phase.
The role of  $t_{pp}$  can be understood
in the strong-coupling limit $U_d, \varepsilon \gg t_{dp}, t_{pp}$ with $U_p=0$.
In that limit one can derive an effective Kondo-lattice model with localized spins 
in $d$-orbitals coupled to doped holes in the lowest 1D band resulting from the 
the hopping terms $t_{pp}$ between $p$-orbitals, as done for the 2D Emery model~\cite{Prelovsek1988,Zaanen1988}.
Localized spins and mobile holes are coupled antiferromagnetically by an exchange coupling
$J \sim t^2_{dp}/(U_d-\epsilon)$.
For $J \ll t_{pp}$ we recover the 1D physics leading to a C1S1 phase at low hole doping.

\begin{figure}
\includegraphics[width=0.48\textwidth]{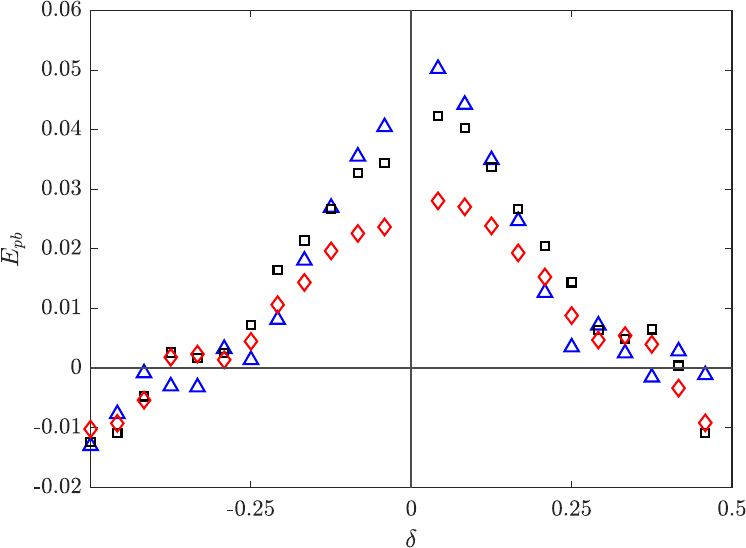}
\caption{\label{fig:pbe5vs3} PBE of two-leg Emery ladders with length $L=24$ as a function of the doping $\delta$
for the second parameter set in Table~\ref{table}.  
Results are for a
three-chain ladder with oxygen sites at its ends (blue triangle), a five-chain ladder with hybridization $t^\prime_{dpy}=t_{dpy}$
(black square),
and a five-chain ladder with hybridization $t^\prime_{dpy}=2t_{dpy}$ (red diamond) between outer oxygen chains and copper sites .
}
\end{figure}

The largest PBE for two doped-holes in a five-chain Emery ladder has been obtained with the
second parameter set listed in Table~\ref{table}. Although we have not investigated the full (five-dimensional) parameter space,
we can confirm two observations made in previous studies.
First, the PBE, when positive, is similar in magnitude to the spin gap.
\changed{Although the relation between PBE and spin gap is not known for the Luther-Emery phase in ladders with repulsive interactions,
it is known that they are equal in the Luther-Emery phase of 1D systems with attractive interaction such as the 1D Hubbard model.}
Second, the largest positive $E_{pb}$ is (at least) one order of magnitude smaller than the other energy scales 
in the system, i.e.  the model parameters and the charge gap of the undoped ladder.

We have found that the properties of the five-chain ladders are qualitatively similar to those of the
three-chain ladders with oxygen sites at their ends.  For instance, Fig.~\ref{fig:pbe5vs3} compares
the PBE of both ladder types as a function of doping. Clearly, we observe a similar dependence on $\delta$.
In addition, we note that the PBE is quite symmetric for hole and electron doping.
In general,
an electron-doped ladder is in a Luther-Emery phase with dominant $d$-wave fluctuations
for a large range of parameters and doping (e.g.,  for the first two parameter sets in Table~\ref{table}).
Hole-doped ladders are often in a Luttinger liquid phase but around the optimal choice of parameters
(i.e., the second parameter set in Table~\ref{table}) the PBE is clearly positive, as seen in Fig.~\ref{fig:pbe5vs3}. In that case
a clear CDW is visible in the hole density variations as \changed{already seen in} three-chain ladders. 
The pair spatial structure~(\ref{eq:pair_structure}) also confirms the presence of bound pairs.
Moreover, pairing correlations are similarly enhanced.
Thus the system is likely in the Luther-Emery phase with dominant $4k_F$-CDW fluctuations but close to the boundary with dominant $d$-wave pairing fluctuations ($K=1/2$),
as described in sec.~\ref{sec:ft}.

Additionally, we have found that removing the ladder-end oxygen sites (or weakening the hybridization
between these sites and the nearest-neighbor copper sites) leads to a localization of the first few doped
electrons at the ladder ends. Conversely, increasing this hybridization leads 
to a localization of the first few doped holes.
Thus the ladder-end oxygen sites have exactly the same effect as in three-chain ladders.
The different behavior of five- and three-chain ladders found in an early work~\cite{Nishimoto2002b}
is therefore entirely due to this boundary effect.

\begin{figure}
\includegraphics[width=0.48\textwidth]{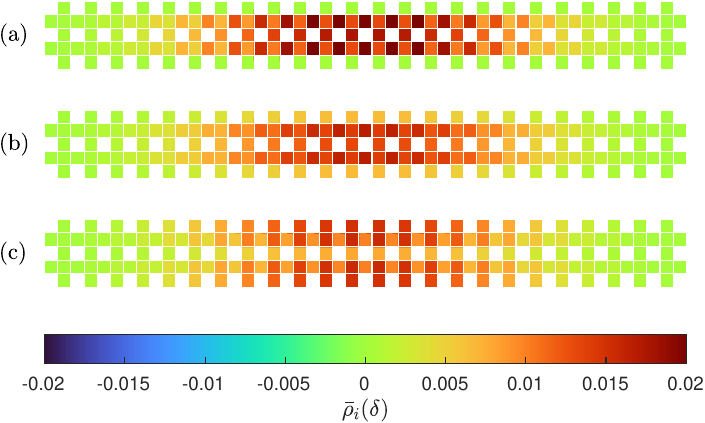}
\caption{\label{fig:density52c} Hole density variation $\bar{\rho}_i(\delta)$  in five-chain ladders with length $L=24$,
two-doped holes ($\delta=1/24$),  and the second parameter set in Table~\ref{table}.  
The hybridization between outer oxygen chains and copper sites is
from top to bottom: (a) $t^\prime_{dpy}=0.01 t_{dpy}$, (b) $t^\prime_{dpy}=t_{dpy}$, and (c) $t^\prime_{dpy}=2t_{dpy}$.
}
\end{figure}

The above discussion raises the question of the influence of the oxygen sites in the two outer chains
(the empty circles in Fig.~\ref{fig:structures}). We have investigated the five-chain ladders with varying
hybridization strength $t_{dpy}^\prime$ between these $p_y$-orbitals and their nearest-neighbor $d$-orbitals.
The calculated properties change smoothly when $t_{dpy}^\prime$ varies from $0$ up to at least $2t_{dpy}$.
The PBE decreases with increasing $t_{dpy}^\prime$ and thus the largest PBE are found 
in three-chain ladders including ladder-end oxygen sites (equivalent to $t_{dpy}^\prime\rightarrow0$ in the five-chain ladder), as shown
in Fig.~\ref{fig:pbe5vs3}.
For hole-doped ladders we also observe a clear change in the hole density variation $\Delta \rho_i(\delta)$
as  $t_{dpy}^\prime$ varies. For weak $t_{dpy}^\prime$ doped holes are mostly located on the 
oxygen sites in the three middle chains while for strong hybridization they are mostly on the
outer oxygen chains, see Fig.~\ref{fig:density52c}. Thus the strength of pairing seems to be related to 
the hole density on the core $p$-orbitals (black circles in Fig.~\ref{fig:structures}).
This observation agrees with our other findings for the pairing properties of hole-doped ladders: the significant weight of the central $p_y$-orbitals in the pair structure and
the strength of the $d-p_y$ pairing correlations~(\ref{eq:correlation_delta2}).
Overall, however, the outer oxygen sites of five-chain ladders do not change the physics
of 2-leg Emery ladders qualitatively, contrary to the ladder-end oxygen sites.

\subsection{Four-chain tubes Cu$_2$O$_4$ \label{sec:4chain}}

Pairing has recently been investigated in the Emery model on a four-chain ladder with periodic boundary conditions in the rung direction~\cite{Jiang2023a,Jiang2023b,Yang2024}.
While these works focus on pair-density-wave quasi-long-range order, the
role of the nearest-neighbor interactions, and long-range hopping, 
they also report a Luttinger liquid (C1S1) phase without
enhanced pairing correlations in the hole-doped system for model parameters comparable to those
used in our work. 

This absence of pairing is due to a non-optimal choice of the Hamiltonian parameters, however.
The most significant problem is the isotropic hopping term between $d$ and $p$-orbitals, $t_{dpy}=t_{dpx}$,
used in~\cite{Jiang2023a,Jiang2023b,Yang2024}.
The lowest two bands of the noninteracting four-chain tube join exactly at this value,
as shown in the appendix. 
Consequently, one expects a transition from the Luther-Emery (C1S0) phase for $t_{dpy} \alt  t_{dpx}$
 to the Luttinger phase (C1S1) for $t_{dpy} \agt t_{dpx}$
 in an interacting system (see the summary of field-theoretical predictions in sec.~\ref{sec:ft}).
In Fig.~\ref{fig:pbe52}  we
show $E_{pb}$ in a ladder with two doped holes as a function of the rung hopping $t_{dpy}$
while other parameters are kept fixed. 
The PBE reaches a maximum around $t_{dpy} \approx 0.7 t_{dpx}$  but is clearly negative
for $t_{dpy} > t_{dpx}$.
The pair spatial structure~(\ref{eq:pair_structure}) also confirms that two doped particles do not build bound pairs
in that regime. For instance, fig.~\ref{fig:pair_structure}(d) shows that the probability of finding the second
doped hole is distributed over the ladder.
 Thus we have to look at the case $t_{dpy} < t_{dpx}$ to find pairing in a system such as the 4-chain Emery tube.
 

\begin{figure}
\includegraphics[width=0.48\textwidth]{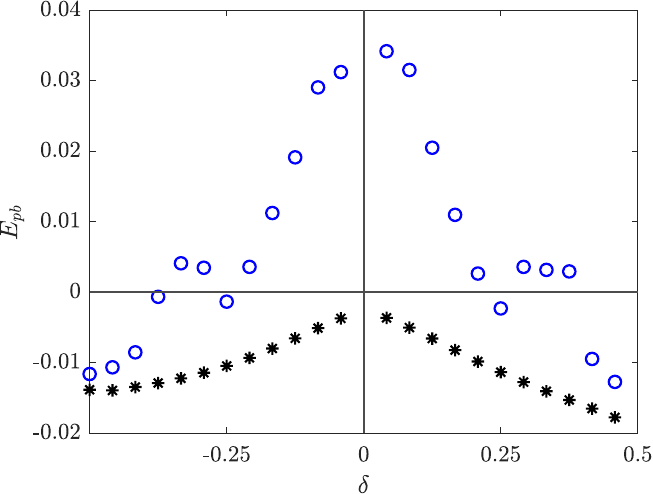}  
\caption{\label{fig:pbe42} PBE of 4-chain Emery tubes with length $L=24$ as a function of the doping $\delta$
for the third parameter set in Table~\ref{table} (blue circle) and for the same parameters but $t_{dpy}=t_{dpx}$ (star).
}
\end{figure}

As \changed{done} for the five-chain ladder we have searched for optimal Hamiltonian parameters
$U_d > 0$, $U_p \geq 0$, $t_{dpy}$, $t_{pp}$, and $\epsilon > 0$
to maximize the PBE of doped holes.  
Note that we have included oxygen sites at 
the tube ends as done for the other ladders to avoid the boundary effects discussed
previously.
The largest PBE for two doped-holes in a four-chain Emery tube has been obtained with parameters
close to the third set listed in Table~\ref{table}.  In particular,  $t_{dpy} \approx 0.7 t_{dpx}$  as seen in Fig.~\ref{fig:pbe52}.

For this optimal parameter set, we again find a Luther-Emery phase with enhanced pairing correlations for a wide doping range.
As an illustration, Fig.~\ref{fig:pbe42}  shows the PBE as a function of doping $\delta$ for the optimal parameters.
Clearly, $E_{pb}$ is positive both for electron and hole doping  up to $\vert \delta \vert \alt 0.2$ and 
is comparable in magnitude to the values found previously in the three and five-chain ladders.
For comparison we also show the PBE for a larger hybridization $t_{dpy}=t_{dpx}$ in Fig.~\ref{fig:pbe42}. Clearly, $E_{pb}$
is always negative in that case.
The existence of bound pairs for the optimal parameters  is confirmed by the analysis of the pair spatial structure~(\ref{eq:pair_structure}).
For instance, fig.~\ref{fig:pair_structure}(e) shows that the probability of finding the second
doped hole is concentrated around the first doped hole, contrary to the case with $t_{dpy}>t_{dpx}$ in fig.~\ref{fig:pair_structure}(d).
Moreover, the analysis of $\bar{S}_i(j)$ confirms again that hole-pairs have a significant weight on the $p_y$-orbitals of the central chain.

In Fig.~\ref{fig:pbe42} we again observe that the PBE behaves similarly for $\delta > 0$ and $\delta < 0$.
Therefore, in all three types of ladders studied here (three-chain and five-chain ladders and four-chain tubes)
the pairing strength appears to be equal for hole and electron doping
if ladder-end oxygen sites are included and optimized model parameters for hole pairing are used.
Nonetheless, the pair structures are different for holes and electrons.

\begin{figure}
\includegraphics[width=0.48\textwidth]{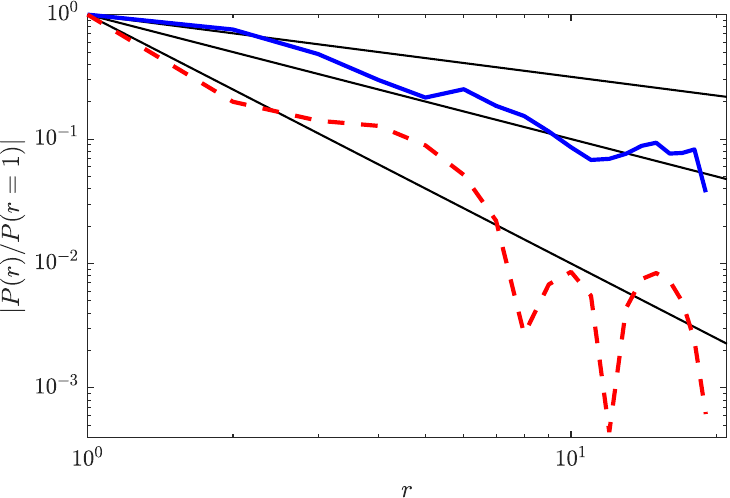}
\caption{\label{fig:correl42} 
Correlation functions for $d$-$p_y$ rung pairs (\ref{eq:correlation_delta2}) in four-chain tubes with length $L=40$ and $\delta = 0.1$
for the second (red dashed) and third  (solid blue)  parameter sets in Table~\ref{table}.
The straight lines correspond to $r^{-\alpha}$ with $\alpha$ = 1/2,1, and 2. }
\end{figure}

As \changed{found in} three-chain and five-chain ladders, we observe that a positive PBE corresponds to
enhanced pairing fluctuations. In particular, 
correlation functions for $d$-$p_y$ rung pairs can decay as slowly as $1/r$ upon hole doping
as shown in Fig.~\ref{fig:correl42}. For comparison, this figure also shows that these pairing correlations
are not enhanced in a four-chain Emery tube with the second parameter set in Table~\ref{table}.
However, we have shown in the previous two sections that this parameter set is optimal for the five-chain ladders resulting in positive PBE and
enhanced pairing in hole-doped (three-chain and five-chain) ladders.
This difference reveals that the parameter values  for which Emery ladders display a Luther-Emery phase
depend significantly on the choice of boundary conditions in the rung direction.
Finally, we also observe a static CDW pinned by the system boundaries in hole-doped four-chain tubes
with positive PBE.

In summary, choosing isotropic parameters ($t_{dpy}=t_{dpx}$) prevents the occurrence of pairing in four-chain Emery tubes.
With a choice  ($t_{dpy}<t_{dpx}$) leading to two bands at the Fermi energy in the noninteracting limit,
we can again obtain results upon hole doping that are consistent with a Luther-Emery phase 
 close to the boundary between dominant $d$-wave pairing and dominant 4$k_F$-CDW
 (i.e., as with a Luttinger exponent $K=1/2$ in the field theory discussed in sec.~\ref{sec:ft}).
 This phase occurs for physically relevant parameters of the Emery model~(\ref{eq:hamiltonian})
 without any of the additional interactions considered in previous works~\cite{Jiang2023a,Jiang2023b,Yang2024}.

\section{Conclusion}

We have investigated the pairing properties of doped Emery ladders with two legs of copper $d$-orbitals
and various numbers of oxygen $p$-orbitals.
Our results show that the signature of pairing is found in the PBE, the spatial structure of doped pairs,
 and in enhanced pairing correlations for hole doping as well as for electron doping with 
realistic parameters for superconducting cuprates.

Ladder systems are often seen as a way towards understanding 2D correlated systems.
However, one should  keep in mind that two-leg ladder systems are still 1D systems
when interpreting their properties. 
In particular, the number of 1D bands at the Fermi energy critically influences 
the low-energy ladder properties at incommensurate band fillings as predicted by field theory.
Our numerical results are compatible with either a Luttinger liquid phase  (C1S1) or a Luther-Emery phase (C1S0)
with competing quasi-long-range orders ($d$-wave pairing and $4k_F$-CDW) depending on the
the parameters and boundary conditions used.
Similarly to previous studies we find that  $d$-wave pairing correlations are dominant in electron-doped ladders.
For hole-doped ladders with optimized but physically reasonable parameters, however, 
 we also find enhanced pairing correlations together with strong static CDW (due to the open boundary conditions), 
 which indicates a Luther-Emery phase with a Luttinger parameter around $K=1/2$.
Our results also show that the phase boundaries and the pairing strength 
depend significantly on model parameters (in particular the ratio between rung and leg hoppings) and boundary conditions.
This explains the absence of Luther-Emery phase or enhanced pairing reported in some previous works.

It is often assumed that the Emery model and the (single-orbital) Hubbard model
describe the same low-energy physics  for parameters and band fillings appropriate for cuprates materials
based on the Zhang-Rice mapping~\cite{Zhang1988}.
The relation between both models and the relevance of the Emery model for doped cuprates have been 
called into question~\cite{Song2021,Song2023} due to the absence of the Luther-Emery phase with dominant $d$-wave pairing correlations
in hole-doped Emery ladders and the asymmetry between electron and hole doping. 
Our results do not support this criticism of the Emery model and its relation to single-orbital Hubbard-like models.

First, on low-energy scales all doped Emery ladders that we have investigated are quite similar to Hubbard two-leg ladders,
in the sense that they belong to the two ``universality classes'',  Luttinger liquid or Luther-Emery phase
for physically relevant parameter regimes.
It is uncontroversial that the pairing properties observed in electron-doped Emery ladders
and those found in Hubbard ladders are strikingly similar.
In contrast, the Luther-Emery phase with competing quasi-long-range orders close to $K=1/2$, 
which is observed in hole-doped Emery ladders, 
is not found in Hubbard ladders with nearest-neighbor hoppings and on-site interactions only. 
However, this simply implies that the Hubbard Hamiltonian must be completed by additional 
interaction terms to reproduce the physics of hole-doped Emery ladders,
such as next-nearest-neighbor interactions~\cite{Giamarchi2003} or hoppings~\cite{Xu2024}.

Second, a quasi-long-range order indicates that the system would like to be in an ordered
state that cannot be stable in 1D but \textit{could} be stable in higher dimensions.
When several quasi-long-range orders compete in the ground state of a 1D system,
one cannot simply assume that the dominant fluctuations (i.e. the correlations with the slowest decay with 
distance) determine the long-range order of the 2D system.
Actually, we have found that for selected parameters hole-doped Emery ladders are in a Luther-Emery phase
with two competing quasi-long-range orders (pairing and CDW).
Although the pairing correlations are not dominant, they are clearly enhanced and thus
could be a precursor of a long-range superconducting order in 2D.
The interplay of pairing and CDW order has been studied theoretically in n-leg $t-J$ and Hubbard ladders 
for several decades~\cite{White1998,Sorella2002,White2003,Hager2005a}.
The question has recently attracted renewed attention~\cite{Jiang2023c,Ponsioen2023,Lu2024,Xu2024,Kaneko2024}.
Several studies report that the superconducting order overcomes the CDW order when 
increasing the number of legs toward a more 2D system~\cite{Jiang2023c,Lu2024}.
Moreover, a recent study of the 2D Emery model reveals a coexistence of charge density modulation and superconductivity~\cite{Ponsioen2023}
using a variational tensor network method
with a supercell that corresponds to the two-leg Emery ladder discussed here.
Thus the observation of enhanced but subdominant pairing correlations in hole-doped Emery ladder
is not evidence against long-rang superconducting order in the 2D hole-doped Emery model.

Third, from a practical point of view, our investigation shows that the PBE is a more reliable quantity than correlation functions
(and derived quantities like the exponent of the power-law decay) for \changed{ascertaining the occurrence of} pairing
in ladder systems using numerical methods such as DMRG.
Accurately determining the Luttinger exponent from the
the power laws~(\ref{eq:dwave}) or~(\ref{eq:cdw})
requires calculating correlation functions over several orders of magnitude
and thus very long ladders. 
\changed{This issue was investigated for Hubbard ladders previously~\cite{Dolfi2015}.}
In contrast, the PBE yields accurate results for shorter ladder lengths.
In particular, 
our PBE results confirm the similar strength of pairing  for hole and electron doping with appropriate model parameters and boundary conditions. 

In conclusion, our study of Emery ladders has not revealed any fact  that rule out the possibility of superconductivity in the hole-doped Emery model on a 2D lattice.
We think that the issue is rather to determine which two-leg ladder structures and model parameters are appropriate to approximate this system.
Indeed, there is an unambiguous two-leg ladder sublattice for the single-orbital Hubbard model and 
the 2D square lattice is obtained by coupling such ladders together.
For the Emery model, however,  different two-leg ladder ladder structures
can be defined depending on the $p$-orbitals taken into account and the boundary conditions in the rung direction.
In this work we have studied three different structures with varying pairing properties.
These three structures possess a reflection symmetry in the rung direction
and a translation invariance  in the leg direction (but for the open boundaries required for DMRG calculations).
Obviously, the 2D cuprate lattice can be recovered if one couples three-chain and five-chain ladders alternatively but not 
by coupling only three-chain ladders or only five-chain ladders together.  
However, one can define several other two-leg ladder structures that breaks reflection symmetry or translation invariance
 but form the Emery model on a 2D cuprate lattice when coupled together.
In addition, one obtains
an isotropic 2D lattice only with equal hopping terms in the leg and rung direction but we have seen that this choice is neither optimal for pairing
nor imposed by rotational symmetry as a ladder model does not have this symmetry anyway.
Therefore, we think studying ladder systems remain a promising approach for understanding superconductivity
in the 2D Emery model but this will require a systematic investigation of the various ladder-like sublattices that 
can be defined from the 2D cuprate lattice.

\acknowledgments
We would like to acknowledge useful discussions with A. Abdelwahab.
GP thanks E.M. Stoudenmire for advice on the use of  iTensor.
The calculations were carried out on the compute cluster at the Leibniz University of Hannover,
which is funded by the Deutsche Forschungsgemeinschaft (DFG, German Research Foundation),
project number  INST 187/742-1 FUGG.

\appendix*

\section{Band structures}

On a periodic 2D lattice the spectrum of the Emery model~(\ref{eq:hamiltonian}) without interaction ($U_p=U_d=0$) 
contains three bands of single-particle eigenstates. Hence this model is often called the three-band Hubbard model.
However, the band structure of the Emery model is quite different on the ladder-like lattices studied in this work.
The non-interacting Hamiltonian~(\ref{eq:hamiltonian}) can be diagonalized exactly for $t_{pp}=0$ and periodic boundary conditions in the leg direction.
In this appendix we assume that $t_{dpx}, t_{dpy} > 0$ and set $\epsilon_p = \epsilon > 0$ as well as $\epsilon_d=0$.
Using the reflection symmetry in the rung direction and the translation invariance in the leg direction,
we obtain
two dispersive even bands of single-particle eigenenergies
\[
E_{\rm e,\pm}(k) = \frac{\epsilon}{2} \pm \sqrt{\left ( \frac{\epsilon}{2} \right )^2 + 2t_{dpy}^2 + [ 2t_{dpx} \cos(ka/2) ]^2 },
\]
two dispersive odd bands
\begin{equation}
\label{eq:disp}
E_{\rm o,\pm}(k) = \frac{\epsilon}{2} \pm \sqrt{\left ( \frac{\epsilon}{2} \right )^2 + [ 2t_{dpx} \cos(ka/2) ]^2 },
\end{equation}
and one flat band at the energy $E=\epsilon$  for the three-chain ladder Cu$_2$O$_3$.
Thus there are 5 bands, as expected for a system with 5 sites per unit cell.

For the five-chain ladder Cu$_2$O$_5$ we find similarly
two dispersive even bands
\[
E_{\rm e,\pm}(k) = \frac{\epsilon}{2} \pm \sqrt{\left ( \frac{\epsilon}{2} \right )^2 + 3t_{dpy}^2 + [ 2t_{dpx} \cos(ka/2) ]^2 },
\]
\\
 two dispersive odd bands
\[
E_{\rm o,\pm}(k) = \frac{\epsilon}{2} \pm \sqrt{\left ( \frac{\epsilon}{2} \right )^2 + t_{dpy}^2 + [ 2t_{dpx} \cos(ka/2) ]^2 },
\]
and three flat bands at $E=\epsilon$.
Thus there are 7 bands, as expected for a system with 7 sites per unit cell.

Finally, for the 4-chain tube Cu$_2$O$_4$ we find 
two dispersive even bands
\[
E_{\rm e,\pm}(k) = \frac{\epsilon}{2} \pm \sqrt{\left ( \frac{\epsilon}{2} \right )^2 + 4t_{dpy}^2 + [ 2t_{dpx} \cos(ka/2) ]^2 },
\]
 two odd bands with the dispersion~(\ref{eq:disp}),
and two flat bands at $E=\epsilon$.
Thus there are 6  bands, as expected for a system with 6 sites per unit cell.

These three different band structures are nonetheless qualitatively similar with respect to the
field-theory predictions for lightly doped ladders in Sec.~\ref{sec:ft}.
In the case of the three-chain and five-chain ladders,
the lowest band $E_{\rm e,-}(k)$ is separated by a band gap from the
second lowest band $E_{\rm o,-}(k)$ for $t_{dpy} > \sqrt{2} t_{dpx}$. 
Thus in the ground state of the undoped ladder 
the band $E_{\rm e,-}(k)$ is filled by holes and the other bands are empty (i.e., filled by electrons).
Upon hole doping the Fermi energy lies in the second lowest band $E_{\rm o,-}(k)$ while it lies in the 
lowest band $E_{\rm e,-}(k)$ band for electron doping.
For $t_{dpy} < \sqrt{2} t_{dpx}$ the bands $E_{\rm e,-}(k)$ and $E_{\rm o,-}(k)$ overlap and 
cross the Fermi energy in the ground state of an undoped or lightly doped ladder.
Thus for weak interactions $U_p$ and $U_d$
we expect a Luttinger liquid (C1S1) phase for $t_{dpy} > \sqrt{2} t_{dpx}$ and a
Luther-Emery (C1S0) phase for $t_{dpy} < \sqrt{2} t_{dpx}$.
For the 4-chain tube, the only difference is that the phase boundary
is at $t_{dpy} = t_{dpx}$. Thus a Luther-Emery phase can only occur 
for anisotropic copper-oxygen hoppings  $t_{dpy} < t_{dpx}$
when periodic boundary conditions are used in the rung direction.

\bibliographystyle{biblev4}
\bibliography{mybibliography}               

\end{document}